\documentclass[final,1p,times]{elsarticle}

\usepackage{amssymb}
\usepackage{graphicx}
\usepackage{bm}

\journal{Nuclear Physics A}

\begin{document}

\begin{frontmatter}

\title{Dineutron correlations and BCS-BEC crossover in nuclear matter with the Gogny pairing force}

\author{Bao Yuan Sun\corref{corr}}\ead{sunby@lzu.edu.cn}
\author{Wei Pan}
\cortext[corr]{Corresponding author}
\address{School of Nuclear Science and Technology, Lanzhou University, 730000 Lanzhou, China}

\begin{abstract}
The dineutron correlations and the crossover from superfluidity of neutron Cooper pairs in the $^1S_0$ pairing channel to Bose-Einstein condensation (BEC) of dineutron pairs in both symmetric and neutron matter are studied within the relativistic Hartree-Bogoliubov theory, with the effective interaction PK1 of the relativistic mean-field approach in the particle-hole channel and the finite-range Gogny force in the particle-particle channel. The influence of the pairing strength on the behaviors of dineutron correlations is investigated. It is found that the neutron pairing gaps at the Fermi surface from three adopted Gogny interactions are smaller at low densities than the one from the bare nucleon-nucleon interaction Bonn-B potential. From the normal (anomalous) density distribution functions and the density correlation function, it is confirmed that a true dineutron BEC state does not appear in nuclear matter. In the cases of the Gogny interactions, the most BEC-like state may appear when the neutron Fermi momentum $k_{Fn}\thicksim0.3~\rm{fm^{-1}}$. Moreover, based on the newly developed criterion for several characteristic quantities within the relativistic framework, the BCS-BEC crossover is supposed to realize in a revised density region with $k_{Fn}\in[0.15,0.63]~\rm{fm^{-1}}$ in nuclear matter.
\end{abstract}


\end{frontmatter}


Pairing correlations are significant phenomena in the fermion systems. For the neutron-neutron pairing, the correlation is supposed to be important in deciding the properties of low-density nuclear matter. One of the evidence comes from the fact that the bare neutron-neutron interaction in the $^1S_0$ channel leads to a virtual state around zero energy characterized by a large negative scattering length $a\approx-18.5\pm0.4~{\rm fm}$ \cite{PRC.36.691}, implying a very strong attraction between two neutrons in the spin singlet state. Furthermore, theoretical predictions suggest that around $1/10$ of the nuclear saturation density $\rho_{0}$, the $^1S_0$ pairing gap may take a remarkably larger value than that around $\rho_0$ \cite{Baldo1990}. In addition, the strong dineutron correlations are also supported by the enhancement of two-neutron transfer cross sections in several heavy nuclei \cite{Oertzen2001}. In the weakly bound neutron-rich nuclei, the couplings to the continuum could strengthen the dineutron correlations, which play a crucial role in unstable nucleus and the formation of the halo \cite{BE1991,Meng2006}. Recently, a ground state dineutron decay was observed for the first time in $^{16}$Be with a small emission angle between the two neutrons, indicating the structure of dineutron clusters inside neutron-rich nuclei \cite{Spyrou}.

The BCS-BEC crossover is one of the hottest issues recently in the study of the pairing correlations in the fermion systems, which is described as a smooth and adiabatic evolution of the pairing from the weakly coupled Bardeen-Cooper-Schrieffer (BCS) type to the strongly correlated Bose-Einstein condensation (BEC) state with increasing pairing strength \cite{Leggett,JLTP1985}.
For the neutron-proton pairing, the BCS-BEC crossover has been investigated in the $^3S_1$-$^3D_1$ channel, in which the strong
spatial correlation and the BEC of the deuterons may occur at low densities \cite{Alm1990,Alm1993,Stein,Baldo1995,Lombardo2001,Sedrakian,HuangXG,MJin2010,Stein2012}. On the other side, the progress in studying the dineutron correlations in weakly bound nuclei has stimulated a lot of interests in searching for possible BCS-BEC crossover of neutron-neutron pairing \cite{Matsuo,Margueron,Isayev,Enyo,Abe1,Abe2,BySun2010,SunTT}. It has been revealed that the dineutron correlations get stronger as density decreases, and the BCS-BEC crossover of the neutron pairing may occur at low densities. The spatial structure of neutron Cooper pair wave function could evolve from BCS-type to BEC-type with decreasing density. In finite nuclei, the coexistence of BCS- and BEC-like spatial structures of neutron pairs has been exhibited in the halo nucleus $^{11}$Li as well \cite{HaginoPRL2007}. From the two-particle wave function, a strong dineutron correlation between the valence neutrons appears on the surface of the nucleus.

As the relativistic mean-field (RMF) theory and its extension to the relativistic Hartree-Bogoliubov (RHB) theory had achieved lots of successes in the descriptions of both nuclear matter and finite nuclei near or far from the stability line \cite{Meng2006,Serot}, the dineutron BCS-BEC crossover in the $^1S_0$ pairing channel was studied within the RHB theory in nuclear matter, taking the bare nucleon-nucleon interaction Bonn-B in the particle-particle ($pp$) channel \cite{BySun2010,SunTT}. It is argued that there is no evidence for a true dineutron BEC state at any density, but some features of the BCS-BEC crossover are seen in the density regions, $0.05~{\rm fm^{-1}}<k_{Fn}<0.7~(0.75)~{\rm fm^{-1}}$, for the symmetric nuclear (pure neutron) matter. In addition, the reference values of several quantities characterizing the BCS-BEC crossover are obtained in the relativistic framework \cite{SunTT}.

Since details of the nucleon-nucleon interaction are still not completely uncovered because of the insensitivity of the experimental data, for convenience, several effective nuclear interactions, such as the zero-range contact force or finite-range Gogny force \cite{Gogny1980}, are widely used in the $pp$ channel in most of investigations on nuclear pairing correlations. Besides, the effective interactions in the RMF theory are also used in the $pp$ channel \cite{Kucharek,LiJun2008}, but one has to introduce an effective factor to this kind of pairing interactions in order to obtain reasonable values for the gap parameter \cite{LiJun2008}. Because the pairing properties strongly depend on the pairing strength, it is expected that the BCS-BEC crossover phenomenon will be influenced by different selections of the pairing interaction. It has been shown that the Gogny interactions predict pairing properties that are surprisingly close to those derived from the bare realistic nuclear interactions \cite{Garrido,Serra}. Recently, a Gogny-Hartree-Fock-Bogoliubov nuclear mass model is presented, which makes great success in reproducing the nuclear masses with an accuracy comparable with the best mass formulas \cite{Hilaire}. Thus, it is of great interest to study the behavior of dineutron correlations and the BCS-BEC crossover in nuclear matter with several Gogny interactions used in the $pp$ channel, and investigate the effects of different pairing forces on the results.

It should be noticed that the bare nucleon-nucleon interaction should be corrected by the medium polarization effects (MPEs), including mainly self-energy contributions and vertex corrections, in extracting the effective interactions in the $pp$ channel at various densities \cite{Lombardo99,Schulze96,Schulze01,Schwenk,FFIS,Cao}. Most of these studies predict a reduction of the pairing gap in neutron matter due to the MPEs. However, it has been suggested that the MPEs enhance the pairing interaction in finite nuclei because of the surface vibrations \cite{BBGVBT}. Encouraged by this apparent contradiction between neutron matter and finite nuclei, a microscopic Brueckner calculation was performed on the MPEs in both symmetric and neutron matters and dramatic results were revealed \cite{Cao}. The MPEs clearly increase the pairing correlations in symmetric nuclear matter (especially for neutron Fermi momentum $k_{Fn}<0.7~{\rm fm^{-1}}$) but decrease them in neutron matter. Further studies claimed that a dineutron BEC state would be formed in symmetric nuclear matter at $k_{Fn}\thicksim0.2~\rm{fm}^{-1}$ if the pairing interaction is screened \cite{Margueron,Isayev}. Moreover, the effects of medium polarization and possible ambiguity of pairing force on the dineutron correlations and the BCS-BEC crossover were investigated in the nuclear matter phenomenologically by taking the bare nucleon-nucleon interaction in the $pp$ channel multiplied by an effective factor \cite{SunTT}. It is shown that if the effective factor is larger than 1.10, a dineutron BEC state appears in the low-density limit, and if it is smaller than 0.85, the neutron Cooper pairs are found totally in the weak coupling BCS region.

In this work, the dineutron correlations in the $^1S_0$ channel for the nuclear matter will be studied within the RHB theory, with three selected finite-range Gogny interactions, i.e., D1 \cite{Gogny1980}, D1S \cite{Gogny1991} and D1N \cite{Chappert2008}, in the $pp$ channel. The evolution of several characteristic quantities for the dineutron correlations, including the normal and anomalous density distribution functions as well as the density correlation function of the neutron Cooper pairs, will be discussed in comparison with the results with Bonn-B potential \cite{BySun2010}, then the BCS-BEC crossover will be investigated with the criteria based on the relativistic calculations \cite{SunTT}. As have been elucidated in the previous study \cite{SunTT}, the MPEs that go beyond the mean-field approximation are not considered in this work. Attention will be paid to the comparison among various versions of the pairing interaction.


In the RHB theory, meson fields are considered dynamically beyond the mean-field theory to provide the
pairing field through the anomalous Green's functions \cite{Kucharek}. For the infinite
nuclear matter, the Dirac-Hartree-Fock-Bogoliubov equation reduces to the usual BCS equation.
For the $^1S_0$ channel, the pairing gap function $\Delta(p)$ is
 \begin{equation}
   \Delta(p) = -\frac{1}{4\pi^2}\int_0^\infty v_{pp}(k,p)\frac{\Delta(k)}{2E_k}k^2dk,
 \end{equation}
where $v_{pp}(k,p)$ is the matrix element of nucleon-nucleon
interaction in the momentum space for the $^1S_0$ pairing channel,
and $E_k$ is the quasi-particle energy. The corresponding normal and anomalous density
distribution function are defined as
 \begin{equation}
   \rho_k = \frac{1}{2} \left[ 1 - \frac{\varepsilon_k - \mu}{E_k} \right],\quad
   \kappa_k = \frac{\Delta(k)}{2E_k},
 \end{equation}
with the single-particle energy $\varepsilon_k$ and the chemical potential $\mu$ obtained from the standard RMF approach.
For nuclear matter with given baryonic density $\rho_b$ and isospin asymmetry $\zeta=(\rho_n-\rho_p)/\rho_b$,
the gap function can be solved by a self-consistent iteration method with no-sea approximation. For more theoretical details we refer the reader to Refs~\cite{Meng2006,BySun2010,SunTT}.

The finite-range Gogny interactions are utilized in the $pp$ channel, which are obtained by combining two Gaussian functions describing a long-range attraction and a short-range repulsion respectively. For the $^1S_0$ pairing channel, the matrix element $v_{pp}(k,p)$ is just the
average of $v_{pp}(\bm{k}, \bm{p})$ over the angle between the vectors $\bm{k}$ and $\bm{p}$, which has the form
\begin{eqnarray}
v_{pp}(k,p)=\frac{4\pi^{3/2}}{kp}\sum_{m=1}^2C_m\exp\left(-\frac{\mu_m^2}{4}(k^2+p^2)\right)\sinh(\frac{1}{2}\mu_m^2kp).
\end{eqnarray}
with the parameters $\mu_m$ and $C_m\equiv\mu_m(W_m-B_m-H_m+M_m)$ taken from the corresponding Gogny interaction \cite{Gogny1980,Gogny1991,Chappert2008}. It is noteworthy that only the density-independent part of the Gogny interaction contributes to the pairing in the $^1S_0$ channel.

In the following studies, the Gogny interactions D1, D1S and D1N will be adopted for $v_{pp}(k,p)$.
For the mean-field calculation in the particle-hole ($ph$) channel, the effective interaction PK1 \cite{Long2004} is used,
since the results do not depend sensitively on various other parameter sets \cite{BySun2010}.
For comparison, the results given by the pairing force with bare nucleon-nucleon interaction Bonn-B \cite{BySun2010} will be discussed again as well.

\begin{figure}
\centering
\includegraphics[width=0.6\textwidth]{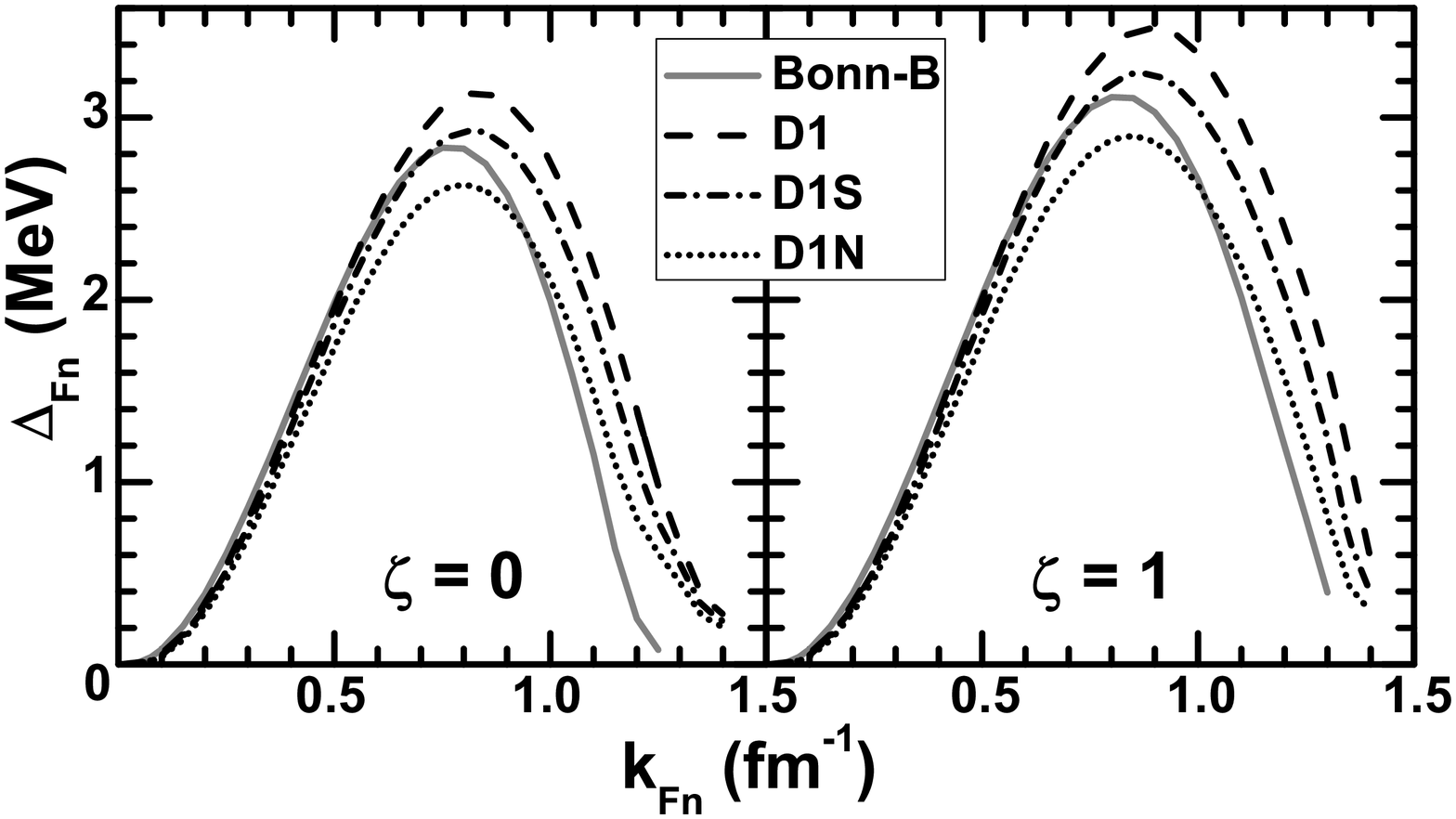}
\caption{Neutron pairing gap at the Fermi surface $\Delta(k_{Fn})$ as a function of the neutron Fermi momentum $k_{Fn}$ in symmetric nuclear matter ($\zeta=0$, left panel) and pure neutron matter ($\zeta=1$, right panel). The effective interaction PK1 \cite{Long2004} is used for the mean-field calculation in the $ph$ channel, and the Gogny interactions D1 (dashed line), D1S (dashed-dotted line) and D1N (dotted line) are used in the $pp$ channel in comparison with Bonn-B potential \cite{BySun2010} (grey solid line).}
\label{fig:Gap}
\end{figure}

To investigate the properties of the neutron Cooper pairs, it is interesting to look into the neutron pairing gap $\Delta(p)$ first. One of the most important properties of the neutron pairing gap is its value at the Fermi surface $\Delta_{Fn}\equiv\Delta(k_{Fn})$. In Fig.~\ref{fig:Gap}, $\Delta_{Fn}$ is shown as a function of the Fermi momentum $k_{Fn}$ for different pairing interactions in both symmetric nuclear matter and pure neutron matter. It is found that the neutron pairing gap $\Delta_{Fn}$ is strongly dependent on the Fermi momentum, or equivalently, the nuclear matter density. $\Delta_{Fn}$ increases as the Fermi momentum (or density) goes down, reaches a maximum at $k_{Fn}\thickapprox0.8~{\rm fm^{-1}}$ in symmetric nuclear matter or $k_{Fn}\thickapprox0.9~{\rm fm^{-1}}$ in pure neutron matter, and then rapidly drops to zero. A systematical enhancement of about 0.3 MeV for $\Delta_{Fn}$ around $k_{Fn}=0.8~{\rm fm^{-1}}$ is revealed in pure neutron matter compared with those in symmetric nuclear matter for all of the adopted pairing interactions.

In comparison with the Bonn-B potential, it is seen that the Gogny interactions especially D1 have larger pairing gap when approaching to the saturation density. At $k_{Fn}=0.8~\rm{fm^{-1}}$, among the adopted pairing interactions, D1 gives the maximum values of $\Delta_{Fn}$, namely, 3.13 MeV in symmetric nuclear matter and 3.40 MeV in pure neutron matter. As the density decreases, the values of $\Delta_{Fn}$ from D1, D1S and D1N become first consistent with and then smaller than the one from Bonn-B potential gradually. As illustrated in a recent work \cite{SunTT}, the pairing strength required for appearance of a dineutron BEC state in the low-density limit must be stronger than 1.1 times of Bonn-B potential, and the corresponding pairing gap $\Delta_{Fn}$ is 4.12 MeV around $k_{Fn}=0.8~\rm{fm^{-1}}$. Therefore, it is expected here again with the Gogny pairing interactions that a true dineutron BEC state cannot occur at any density in nuclear matter.

\begin{figure}
\centering
\includegraphics[width=0.5\textwidth]{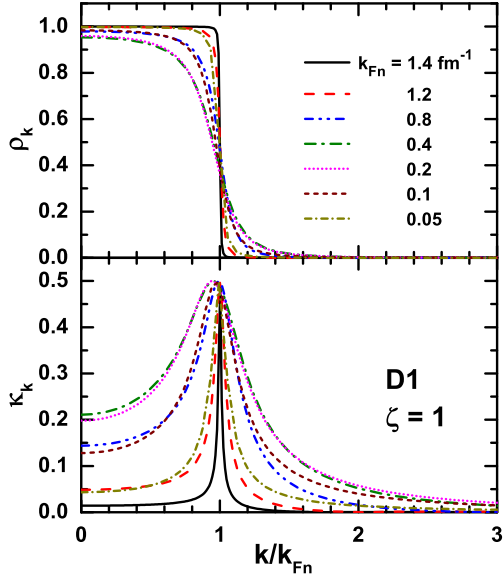}
\caption{Neutron normal and anomalous density distribution function, $\rho_k$ and $\kappa_k$, as a function of the ratio of the neutron momentum to the Fermi momentum $k/k_{Fn}$ at several neutron Fermi momenta $k_{Fn}$ in pure neutron matter. The effective interaction PK1 \cite{Long2004} is used for the mean-field calculation in the $ph$ channel and the Gogny interaction D1 is used in the $pp$ channel.}
\label{fig:v2uv}
\end{figure}

The normal and anomalous density distribution functions, $\rho_k$ and $\kappa_k$, provide us valuable informations on the mean field and the pairing field. In Fig.~\ref{fig:v2uv}, neutron normal and anomalous density distribution functions are plotted as a function of $k/k_{Fn}$ at several neutron Fermi momenta $k_{Fn}$ in pure neutron matter with the Gogny pairing interaction D1. When $k_{Fn}=1.4~\rm{fm^{-1}}$, the normal density distribution function $\rho_k$ is represented nearly as a step function at $k/k_{Fn}=1.0$, and the neutron Cooper pair could be treated as a BCS-like pair. With decreasing Fermi momentum, the momentum distribution of $\rho_k$ evolves smoothly and deviates from the step function gradually. After reaching a maximum deviation from the step function at $k_{Fn}=0.2~\rm{fm^{-1}}$ and $0.4~\rm{fm^{-1}}$, the momentum distribution of $\rho_k$ approaches to the step function again at dilute density.

The anomalous density distribution function $\kappa_k$, known as the Cooper pair wave function in momentum space, presents a similar density-dependent behavior to $\rho_k$ as well. When $k_{Fn}=1.4~\rm{fm^{-1}}$, it is seen that the pairing exists almost on the Fermi surface with a well-developed peak of $\kappa_k$. As the density decreases, the momentum distribution of $\kappa_k$ is expanded to both lower and higher regions than the Fermi momentum. After generating a maximum momentum dispersion of $\kappa_k$ at $k_{Fn}=0.2~\rm{fm^{-1}}$ and $0.4~\rm{fm^{-1}}$, the momentum distribution of $\kappa_k$ is narrowed again in the vicinity of the Fermi momentum at dilute area. Therefore, from the evolution of $\rho_k$ and $\kappa_k$ with the Fermi momentum, the most dineutron BEC-like state may appear at the density region of $k_{Fn}\in[0.2,0.4]~\rm{fm^{-1}}$ in the case of the Gogny pairing interaction D1. Similar results are obtained with the other two Gogny pairing interactions D1S and D1N, and for symmetric nuclear matter as well.

As a quantitative study, the density correlation function developed recently for ultracold atomic gases has been suggested as a clear distinction between the BCS and BEC limits, which gives prominence to the relative strength between the mean field and the pairing field \cite{Isayev,BySun2010,Mihaila}. It is expressed as
\begin{equation}
D(q)=I_{\kappa}(q)-I_{\rho}(q).
\end{equation}
At zero-momentum transfer $q=0$, the normal and anomalous density contributions $I_{\rho}$ and $I_{\kappa}$ respectively read
\begin{eqnarray}
I_{\rho}(0)&=&\frac{1}{\pi^2\rho_n}\int_0^{\infty}\rho_k^2k^2dk,\\
I_{\kappa}(0)&=&\frac{1}{\pi^2\rho_n}\int_0^{\infty}\kappa_k^2k^2dk,
\end{eqnarray}
and they satisfy the sum rule $I_{\rho}(0)+I_{\kappa}(0)=1$. The sign change of the density correlation function at zero-momentum transfer $D(0)$ could be taken as a criterion of the BCS-BEC crossover \cite{Mihaila}, i.e., $D(0)<0$ means a BCS-type pairing and $D(0)>0$ represents a dineutron BEC state.

\begin{figure}
\centering
\includegraphics[width=0.6\textwidth]{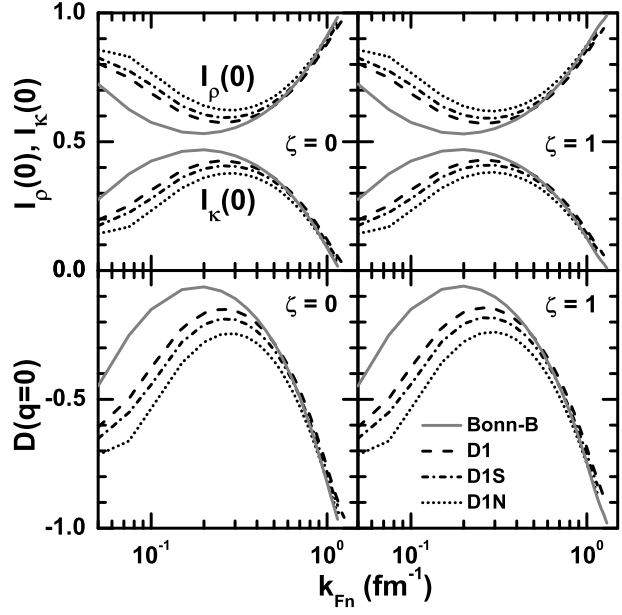}
\caption{Zero-momentum transfer density correlation function $D(q=0)$ and its normal and anomalous density contributions, $I_{\rho}(0)$ and $I_{\kappa}(0)$, as a function of the neutron Fermi momentum $k_{Fn}$ in symmetric nuclear matter ($\zeta=0$, left panels) and pure neutron matter ($\zeta=1$, right panels). The effective interaction PK1 \cite{Long2004} is used for the mean-field calculation in the $ph$ channel, and the Gogny interactions D1 (dashed line), D1S (dashed-dotted line) and D1N (dotted line) are used in the $pp$ channel in comparison with Bonn-B potential \cite{BySun2010} (grey solid line).}
\label{fig:DCF}
\end{figure}

In Fig.\ref{fig:DCF}, the density correlation functions at zero-momentum transfer $D(0)$ and its normal and anomalous density contributions, $I_{\rho}(0)$ and $I_{\kappa}(0)$, are shown as a function of $k_{Fn}$ in nuclear matter. In three cases of Gogny interactions, as the density decreases, it is seen that $I_{\rho}(0)$ first drops down from 1.0, reaches a minimum at $k_{Fn}\thicksim$0.3~fm$^{-1}$, and then goes up again, while $I_{\kappa}(0)$ gives opposite trend and has no contribution when approaching to the saturation density. Thus, a peak for the density correlation function $D(0)$ is found around $k_{Fn}=0.3~{\rm fm^{-1}}$ for the Gogny interactions, larger than the value of $k_{Fn}=0.2~{\rm fm^{-1}}$ in the case of the Bonn-B potential. Therefore, the most BEC-like state may appear around $k_{Fn}=0.3~{\rm fm^{-1}}$ in the calculations with the Gogny interactions, which is in agreement with the former prediction by the normal and anomalous density distribution functions, but distinguished from the Bonn-B case \cite{BySun2010}. Because the anomalous density contribution $I_{\kappa}(0)$ is always smaller than the normal one $I_{\rho}(0)$, the density correlation function $D(0)$ never changes sign at all for all of adopted pairing interactions. According to the criterion mentioned above, dineutrons are not in BEC state but just in the transition region between BCS and BEC regimes. Compared with the result of Bonn-B potential, the values of $D(0)$ are smaller in the cases of three Gogny interactions so that the corresponding dineutron behavior is farther away from the BEC limit. A further verification is performed here by the quasi-particle excitation spectrum with the similar method discussed in Ref. \cite{BySun2010}, and the consistent conclusion that there is no evidence for the appearance of a true dineutron BEC state at any density is gained.

\begin{figure}
\centering
\includegraphics[width=0.6\textwidth]{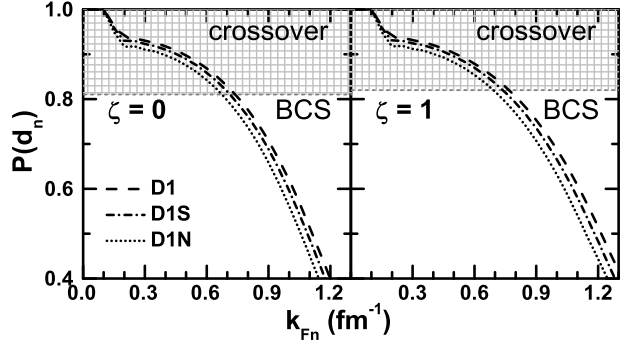}
\caption{Probability $P(r)$ for the partner neutrons correlated within the average inter-neutron distance $r=d_n$ as a function of the neutron Fermi momentum $k_{Fn}$ in symmetric nuclear matter ($\zeta=0$, left panel) and pure neutron matter ($\zeta=1$, right panel). The effective interaction PK1 \cite{Long2004} is used for the mean-field calculation in the $ph$ channel, and the Gogny interactions D1 (dashed line), D1S (dashed-dotted line) and D1N (dotted line) are used in the $pp$ channel. The referred BCS-BEC crossover regions \cite{SunTT} (grey grid) are given as well.}
\label{fig:Pdn}
\end{figure}

As discussed above, the Gogny pairing interactions have lead to several recognizable differences from the Bonn-B potential for the neutron pairing properties in nuclear matter. Hence, it is desirable to investigate the influence of the pairing interaction on features of the BCS-BEC crossover. Lots of characteristic quantities have been introduced to describe the BCS-BEC crossover quantitatively \cite{Matsuo,Margueron,BySun2010,PRL.71.3202}. The probability $P(r)$ for the pair partners within a relative distance $r$ is such a quantity to clarify the spatial correlation of the neutron Cooper pairs, which is defined as
\begin{equation}
P(r)=\int_0^r|\Psi_{pair}(r')|^2r'^2 dr',
\end{equation}
where $\Psi_{\rm pair}(r)$ is the wave function of the neutron Cooper pairs in coordinate space,
 \begin{equation}
 \Psi_{\rm pair}(r) = \frac{C}{(2\pi)^3}\int
 \kappa_ke^{i\bm{k}\cdot\bm{r}}d\bm{k},
 \end{equation}
with the constant $C$ determined from the normalization condition.

In Fig.\ref{fig:Pdn}, the probability $P(d_n)$, i.e., $P(r)$ for the pair partners within the average inter-neutron distance $d_n\equiv\rho_n^{-1/3}$ is shown as a function of $k_{Fn}$ with the different Gogny pairing interactions in both symmetric nuclear matter and pure neutron matter. It is revealed that the values of $P(d_n)$ increase monotonically with decreasing density, and approach to 1.0 at dilute area. From the newly developed criterion based on the RHB theory, it is suggested that the BCS-BEC crossover is determined by $P(d_n)>0.81$ in symmetric nuclear matter and $P(d_n)>0.82$ in pure neutron matter respectively \cite{SunTT}. This limits the Fermi momentum for the realization of the BCS-BEC crossover in $k_{Fn}<0.75~(0.78)~\rm{fm^{-1}}$ (parameter set D1), $k_{Fn}<0.70~(0.75)~\rm{fm^{-1}}$ (parameter set D1S), and $k_{Fn}<0.65~(0.70)~\rm{fm^{-1}}$ (parameter set D1N) in symmetric nuclear (pure neutron) matter, as shown in the columns marked with UL$^\ast$ in Table~\ref{tab}.

\begin{figure}[h]
\centering
\includegraphics[width=0.6\textwidth]{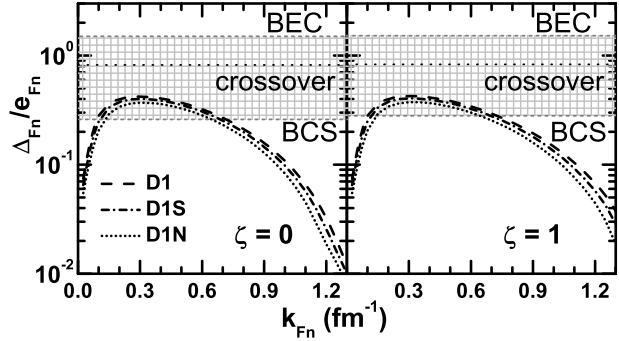}
\caption{The ratio $\Delta_{Fn}/e_{Fn}$ between the neutron pairing gap at the Fermi surface $\Delta_{Fn}$ and the neutron Fermi kinetic energy $e_{Fn}$ as a function of the neutron Fermi momentum $k_{Fn}$ in symmetric nuclear matter ($\zeta=0$, left panel) and pure neutron matter ($\zeta=1$, right panel). The effective interaction PK1 \cite{Long2004} is used for the mean-field calculation in the $ph$ channel, and the Gogny interactions D1 (dashed line), D1S (dashed-dotted line) and D1N (dotted line) are used in the $pp$ channel. The referred BCS-BEC crossover regions (grey grid) and the unitary limits (dotted line) \cite{SunTT} are given as well.}
\label{fig:Del}
\end{figure}

The BCS-BEC crossover could be explored as well by the ratio $\Delta_{Fn}/e_{Fn}$ between the neutron pairing gap at the Fermi surface $\Delta_{Fn}$ and the neutron Fermi kinetic energy $e_{Fn}\equiv e(k_{Fn})$. Here the neutron kinetic energy $e(k)$ is given by
 \begin{equation}
 e(k)=\sqrt{k^2 + M^{\ast2}} -M^\ast,
 \end{equation}
where the scalar mass $M^\ast$ can be obtained from the standard RMF approach \cite{Meng2006}. If the ratio is large enough, the neutron is expected to be in the strong coupling regime. From the criterion based on the RHB theory, the BCS-BEC crossover region is estimated as $0.26\lesssim\Delta_{Fn}/e_{Fn}\lesssim1.50$ in symmetric nuclear matter and $0.28\lesssim\Delta_{Fn}/e_{Fn}\lesssim1.52$ in pure neutron matter, while the unitary limit is given by $\Delta_{Fn}/e_{Fn}\thickapprox0.82~(0.83)$ in symmetric nuclear (pure neutron) matter. One should notice that the values of the criterion given here are signally different from those obtained in the regularized contact interaction model of non-relativistic framework \cite{Matsuo,Margueron}. Therefore, the density region for the realization of BCS-BEC crossover is supposed to change more or less compared to the previous results \cite{BySun2010}, even determined differently for symmetric nuclear matter and pure neutron matter.

In Fig.\ref{fig:Del}, the ratios $\Delta_{Fn}/e_{Fn}$ are plotted as a function of the neutron Fermi momentum in nuclear matter based on three adopted Gogny interactions. With decreasing Fermi momentum, the values grow up first and go into BCS-BEC crossover region. It is found that the curves never intersect with the line of the unitary limit, i.e., the dineutron cannot perform a BEC state. At very low densities, the curves bend downward and return back to the BCS region again. Estimated from the criterion of $\Delta_{Fn}/e_{Fn}$ referred above, the BCS-BEC crossover is marked in the density region with $0.08~(0.10)~{\rm fm^{-1}}<k_{Fn}<0.72~(0.73)~{\rm fm^{-1}}$ (parameter set D1), $0.10~(0.12)~{\rm fm^{-1}}<k_{Fn}<0.70~(0.70)~{\rm fm^{-1}}$ (parameter set D1S), and $0.14~(0.15)~{\rm fm^{-1}}<k_{Fn}<0.63~(0.63)~{\rm fm^{-1}}$ (parameter set D1N) respectively for symmetric nuclear (pure neutron) matter. The results are summarized in Table~\ref{tab}, where the values determined from the newly developed criteria are also listed for the bare nucleon-nucleon interaction Bonn-B.

From Table~\ref{tab}, it is shown that the density region for the realization of the BCS-BEC crossover is affected clearly by the pairing force. In fact, the Gogny interaction D1N gives the weakest pairing gap at low densities as revealed in Fig. \ref{fig:Gap}, so it generates a narrowest range of density for the BCS-BEC crossover, namely, $0.14 (0.15)~{\rm fm^{-1}}<k_{Fn}<0.63~{\rm fm^{-1}}$ for symmetric nuclear (pure neutron) matter. Therefore, after combining the above analyses from the probability $P(d_n)$ and the ratio $\Delta_{Fn}/e_{Fn}$, the BCS-BEC crossover is supposed to appear most likely in the density region with $k_{Fn}\in[0.15,0.63]~\rm{fm^{-1}}$ in nuclear matter, supported by not only three effective Gogny interactions but also bare nucleon-nucleon interaction Bonn-B. Additionally, it should be announced that if one utilizes the criterion obtained from the regularized contact interaction model in the non-relativistic framework \cite{Matsuo,Margueron}, the BCS-BEC crossover region will turn out to be $k_{Fn}\in[0.11,0.72]~\rm{fm^{-1}}$, a little broader than the one from newly developed criteria just mentioned.

\begin{table}
\caption{The lower (LL) and upper limits (UL) of the neutron Fermi momentum $k_{Fn}$ for the realization of BCS-BEC crossover determined by the referred criteria of the ratio $\Delta_{Fn}/e_{Fn}$, and the upper limits (UL$^\ast$) determined by the referred criteria of the probability $P(d_n)$ \cite{SunTT}, with different pairing interactions in symmetric nuclear matter ($\zeta=0$) and pure neutron matter ($\zeta=1$). The values have units of fm$^{-1}$.}
\label{tab}
\centering
\begin{tabular}{lcccccc}
\hline\hline
& \multicolumn{3}{c}{$\zeta=0$} & \multicolumn{3}{c}{$\zeta=1$}\\
\cline{2-4\ \ \ }\cline{5-7\ } & LL & UL & UL$^\ast$ & LL & UL & UL$^\ast$\\
\hline
D1     & 0.08 & 0.72 & 0.75 & 0.10 & 0.73 & 0.78\\
D1S    & 0.10 & 0.70 & 0.70 & 0.12 & 0.70 & 0.75\\
D1N    & 0.14 & 0.63 & 0.65 & 0.15 & 0.63 & 0.70\\
Bonn-B & 0.05 & 0.70 & 0.70 & 0.05 & 0.70 & 0.75\\
\hline\hline
\end{tabular}
\end{table}

As a useful supplement, it is worthwhile to have a discussion about the role of the scattering phase shifts in the pairing properties. Several commonly used bare or effective nuclear interactions in the $pp$ channel are deduced virtually from fitting the measured data of scattering phase shifts \cite{Margueron,EBH1997}. Thus, corresponding pairing properties obtained are in agreement with those from quantum Monte Carlo calculations more or less, such as in Ref. \cite{FFIS}. Several efforts have been made to give a direct analysis of the dineutron correlations in the low-density region from the scattering data \cite{Matsuo,Margueron,SunTT}, where the scattering length $a$, defined in terms of the T-matrix for the scattering in the free space, was used. According to the definition, the negative value of the scattering length $a$ indicates an unbound state of neutron Cooper pairs while the positive value represents a bound state. It has been proposed to define the boundaries of the BCS-BEC crossover using a regularized gap equation approach~\cite{Leggett,Matsuo,SunTT,PRL.71.3202}, which is related with the scattering length $a$. It was shown that the properties of pairing correlations can be uniquely controlled by a dimensionless parameter $1/(k_{Fn}a)$ which can give the evolution from BCS to BEC. The dimensionless parameter $1/(k_{Fn}a)\ll-1$ corresponds to the weak coupling BCS regime, while $1/(k_{Fn}a)\gg1$ is related to the strong correlated BEC regime. The boundaries characterizing the BCS-BEC crossover can be approximately determined by $1/(k_{Fn}a)=\pm1$, and the unitary limit is defined as $1/(k_{Fn}a)=0$, which is the midpoint of the BCS-BEC crossover. With this method, the BCS-BEC crossover region of the neutron pairing was determined within different many-body theories \cite{Matsuo,Margueron,SunTT}. Similar investigation is performed here with three selected Gogny pairing interactions as well, and the results show good consistence with above analysis of the ratio $\Delta_{Fn}/e_{Fn}$.

However, one should be aware that some approximations have been introduced into the calculations of the regularized gap equation from the current RHB theory. The regularized gap equation approach is generically applied to dilute systems, where the interaction matrix elements $v_{pp}(k,p)$ can be approximately treated as constant $v_{0}$, while in the RHB calculations $v_{pp}(k,p)$ and correspondingly the pairing gap $\Delta(k)$ are found to be momentum dependent. Thus, such regularized approach gives poor knowledge on pairing properties at high densities. On the other hand, at very low densities the mean-field approach seems to be inefficient and some new features of nuclear matter such as clustering should be considered \cite{PRL.104.202501,Typel2010}. More sophisticated approaches for the low-density nuclear matter is from the quantum Monte Carlo techniques \cite{Abe1,Abe2,FFIS,MC2005,MC2009} or from the virial expansion method \cite{VirialA,VirialB}, which are the best suited tools for treating strongly correlated systems thus could give reliable discussion on the large scattering length physics. By comparing the curves of the normal density distribution function $\rho_k$ in Fig.~\ref{fig:v2uv} with those from quantum Monte Carlo calculations as shown in Ref. \cite{MC2005}, one find the results in Fig.~\ref{fig:v2uv} are located between the case of the step function and of the unitary limit, which confirm again that dineutrons at low densities are not in BEC state but just in the transition region between BCS and BEC regimes. In the virial expansion method, the second virial coefficient is related to the nucleon-nucleon scattering phase shifts so that the equation of state of low-density nuclear matter can be derived directly from the measured scattering data. Therefore, it is deserved to take a look at the pairing properties with such kind of method.


In conclusion, the dineutron correlations and the BCS-BEC crossover phenomenon for nuclear matter in the $^1S_0$ channel have been investigated based on the relativistic Hartree-Bogoliubov theory with the effective RMF interaction PK1 in the $ph$ channel and the finite-range Gogny force in the $pp$ channel. The influence of the pairing strength on the behaviors of dineutron correlations is performed by comparing the results given by three selected Gogny interactions, i.e., D1, D1S and D1N, and bare nucleon-nucleon interaction Bonn-B as well. It is found that the neutron pairing gaps at the Fermi surface $\Delta_{Fn}$ from three adopted Gogny interactions are smaller than those from the Bonn-B potential at low nuclear densities. From the normal and anomalous density distribution functions, $\rho_k$ and $\kappa_k$, and the density correlation function $D(0)$, it is confirmed that a true dineutron BEC state does not occur at any density in nuclear matter, and neutron pairing is just in the region of BCS-BEC crossover at low densities, in agreement with our previous studies \cite{BySun2010,SunTT}. However, from the calculations with the Gogny interactions, the most BEC-like state may appear at $k_{Fn}\thicksim0.3~\rm{fm^{-1}}$, which is larger than the value of $k_{Fn}\thicksim0.2~\rm{fm^{-1}}$ in the case of the Bonn-B potential. Moreover, based on the newly developed criteria for the characteristic quantities \cite{SunTT}, namely, the probability $P(d_n)$ and the ratio $\Delta_{Fn}/e_{Fn}$, the BCS-BEC crossover is marked in the density region with $k_{Fn}\in[0.15,0.63]~\rm{fm^{-1}}$ in nuclear matter, showing a narrower and more believable range than $k_{Fn}\in[0.11,0.72]~\rm{fm^{-1}}$ determined from the regularized contact interaction model in the non-relativistic framework and $k_{Fn}\in[0.05,0.70]~\rm{fm^{-1}}$ in our previous study as well \cite{BySun2010}.

\section*{Acknowledgements}
This work is partly supported by the National Natural Science Foundation of China (Grant No. 11205075),
and the Fundamental Research Funds for the Central Universities (Grant Nos. lzujbky-2012-k07 and lzujbky-2012-7).


\end{document}